\documentclass[conference,final]{IEEEtran}

\usepackage{cite}
\usepackage{amsmath,amssymb,amsfonts, amsthm}
\usepackage[table]{xcolor}
\usepackage{graphicx}
\usepackage{textcomp}
\usepackage{graphicx}
\usepackage{multirow}
\usepackage[ruled,linesnumbered,vlined]{algorithm2e}
\usepackage{algpseudocode}
\usepackage{comment}
\usepackage{multicol}
\usepackage{vwcol}
\usepackage{tikz}
\usepackage[most]{tcolorbox}
\usepackage{listings}
\usepackage[hidelinks]{hyperref}
\usepackage{xspace}

\usepackage{ifdraft}
\ifoptionfinal{\usepackage[final]{fixme}}{\usepackage[draft]{fixme}}
\fxsetup{mode=multiuser,theme=color,layout=inline,nomargin} 
\FXRegisterAuthor{DL}{dl}{\color{teal}\fbox{DL}}
\FXRegisterAuthor{KR}{kr}{\color{blue}\fbox{KR}}
\FXRegisterAuthor{BW}{bw}{\color{olive!80!black}\fbox{BW}}
\FXRegisterAuthor{RB}{rb}{\color{purple}\fbox{RB}}
\FXRegisterAuthor{NMG}{nmg}{\color{red!90!black}\fbox{NMG}}

\usepackage{booktabs}  
\usepackage{tabularx}  

\makeatletter
\let\MYcaption\@makecaption
\makeatother
\usepackage[font=footnotesize,format=hang,justification=RaggedRight,skip=0pt,belowskip=-.5em]{subcaption}

\makeatletter
\let\@makecaption\MYcaption
\makeatother

\usetikzlibrary{fit}
\usetikzlibrary{patterns}

\usepackage{enumitem}
\setlist[itemize]{leftmargin=*}
\setlist[enumerate]{leftmargin=*}
\newcommand{\qlist}[1]{\par\noindent\rule{0pt}{1.0\baselineskip}\textbf{#1}}   
\newcommand{\dlist}[1]{\qlist{#1:}}

\SetCommentSty{commentfont}
\SetAlFnt{\footnotesize}
\SetAlCapFnt{\footnotesize}
\SetAlCapNameFnt{\footnotesize}

\newcommand{\ml}{ML\xspace}

\begin{document}

\title{Statistical Modeling and Uncertainty Estimation of LLM Inference Systems}

\author{}

{}

\author{
\IEEEauthorblockN{
Kaustabha Ray$^{\dagger}$, Nelson Mimura Gonzalez$^{+}$, Bruno Wassermann$^{\ddagger}$,
Rachel Tzoref-Brill$^{\ddagger}$, Dean H. Lorenz$^{\ddagger}$
}
\IEEEauthorblockA{
$^{\dagger}$IBM Research – India \quad
$^{+}$IBM TJ Watson Research Center \quad
$^{\ddagger}$IBM Research – Israel
}
}

\maketitle

\begin{abstract}
  Large Language Model (LLM) inference systems present significant challenges in statistical performance characterization due to dynamic workload variations, diverse hardware architectures, and complex interactions between model size, batch processing, and throughput requirements. Accurate statistical characterization enables better workload scheduling, adaptive resource provisioning, and cost-aware inference optimization, making it crucial for improving efficiency in large-scale AI deployments. Traditional analytical models provide explainability but cannot cover the vast diversity of real-world workloads, making it impossible to benchmark every scenario in advance. Machine learning (ML) approaches effectively predict performance for non-benchmarked cases but struggle when extrapolating beyond their observed training space. To address these limitations for LLM inference systems, we propose an Analytical with Learning Augmentation (ALA) framework that bridges analytical modeling with \ml for robust statistical prediction and uncertainty estimation in LLM inference workloads.
  Our method employs an analytical throughput model with parameters estimated for benchmarked workloads, then extends to unobserved configurations using \ml predictions. We enhance this with simulated annealing to exploit subsets of the workload data point combinations and develop an error predictor. Finally, we quantify uncertainty based on vector space similarity between new and observed workloads to ensure robust generalization.
Through extensive experimentation on diverse LLM inference workloads, we demonstrate that our framework achieves low median errors while maintaining adaptability to new inference scenarios. The proposed ALA framework provides a principled approach to balancing model fidelity, generalization, and confidence estimation, offering a scalable solution for optimizing LLM inference statistically across cloud and edge environments.
\end{abstract}


\IEEEpeerreviewmaketitle

\section{Introduction} \label{sec:introduction}

\noindent
Large Language Models (LLMs) have become a transformative Artificial Intelligence (AI) innovation, significantly advancing Natural Language Processing (NLP) and text generation. Models such as GPT \cite{brown2020language}, LLaMA \cite{dubey2024llama3herdmodels}, and LaMDA \cite{thoppilan2022lamda} have gained widespread recognition for their ability to comprehend and generate human-like text across diverse tasks. These models are now integral to applications like content creation, question-answering, and language translation, spanning multiple domains, including scientific \ml. The landscape of LLMs has undergone rapid transformation in recent years, driven by three key trends: the rise of open-source LLMs, advancements in specialized AI accelerators, and the development of efficient inference frameworks. These advancements collectively aim to improve LLM performance by leveraging optimized hardware and software solutions. 

A fundamental pillar of LLMs is the inference system \cite{chitty2023survey, zhou2024survey, park2024inference}, which orchestrates the end-to-end process of transforming user input queries into generated content with efficiency and accuracy.  Efficient inference is fundamental to powering LLM driven applications. However, as LLMs continue to scale in size and complexity, optimizing inference becomes increasingly important to ensure a balance between computational efficiency, energy consumption, response latency, and the monetary cost incurred by different workload configurations. 
Statistical predictive models quantify inference systems, capturing the intricate relationships between the input/output load and the resulting performance characteristics such as throughput, latency and so on.  These predictive models are used to drive optimization platforms towards identifying bottlenecks, and explore trade-offs towards LLM inference optimization strategies, such as model placement, request routing, and hardware-aware scheduling. Accurate statistical characterization is thus a critical endeavour.

Predictive models for LLM inference face challenges from the heterogeneity of LLM models and hardware as well as the non-linear scaling behavior of statistical performance measures with workload characteristics. Analytical models provide insights into the relationships between various system parameters and statistical performance metrics, but they often struggle to capture the complexity of real-world workloads. \ml-based approaches, on the other hand, can learn from historical data and make accurate predictions, but they may lack interpretability and generalization capabilities that tend to reduce their accuracy when attempting to extrapolate well outside the range of previously encountered data.

In this paper, we introduce the Analytical with Learning Augmentation (ALA) framework, designed to accurately and robustly predict the statistical characteristics of LLM inference systems. ALA addresses the limitations of traditional analytical modeling and \ml approaches by integrating both methods. It provides predictions of statistical performance characteristics and quantifies uncertainty for unobserved LLM workloads. ALA is capable of making predictions within the range of values observed during benchmarking and extrapolating beyond this range, while also providing an uncertainty measure for these predictions. We discuss and empirically evaluate ALA’s effectiveness in predicting LLM inference throughput.
The main contributions of our paper are as follows: 

  \dlist{%
  Analytical with Learning Augmentation (ALA) framework} 
  We introduce the ALA framework, which combines traditional analytical modeling with \ml techniques. This novel approach extends the applicability of analytical throughput models using learning-based models to predict model parameters for unseen workload configurations. Additionally, ALA employs simulated annealing to systematically navigate the search space of training point combinations to train an error predictor. This enables precise estimation of inference performance prediction accuracy. ALA also performs uncertainty quantification for previously unobserved workloads by analyzing vector space similarity between new workloads and historical training data, ensuring robust generalization.
  \dlist{%
  Empirical evaluation of ALA} Through comprehensive experiments on diverse LLM inference workloads, we demonstrate that our framework achieves low errors while maintaining adaptability to new inference scenarios, offering a scalable solution for optimizing LLM inference performance statistics.

The rest of this paper is organized as follows:  Our methodology is described in Section~\ref{sec:methodology} and evaluated in Section~\ref{sec:experiments}. Related work is discussed in Section~\ref{sec:background}. Finally, in Section~\ref{sec:conclusion} we summarize our contributions  and outline future work.

\section{Methodology}
\label{sec:methodology}

\begin{figure}
        \includegraphics[scale=0.55]{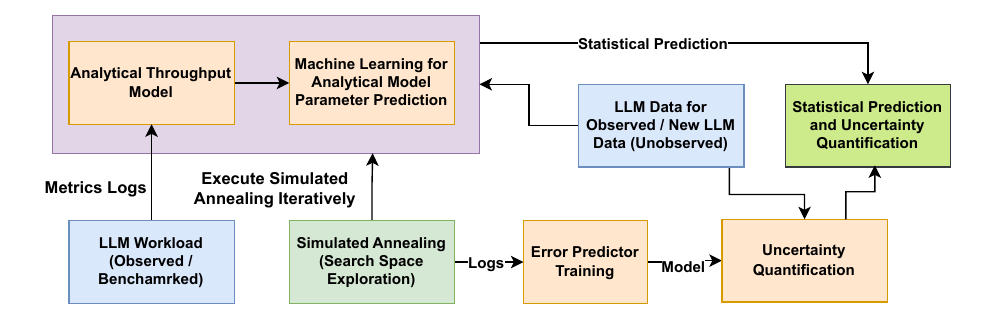}
    \caption{ALA Framework Solution Overview} 
    \label{fig:solarch}
\end{figure}

\noindent
The Analytical with Learning Augmentation (ALA) framework, depicted in Fig.~\ref{fig:solarch}, provides comprehensive predictions of statistical performance characteristics for unobserved LLM workloads through a hybrid approach that combines analytical modeling with \ml techniques. 

  \dlist{%
  Initial LLM workload observations from benchmarking} The first input to ALA are performance data for LLM workloads from benchmarking. These initial data points serve as the cornerstone for the analytical model and the \ml aspects.
  \dlist{%
  Analytical model construction} Using this initial dataset, ALA constructs an analytical throughput model. The framework methodically estimates key parameters from the observed data points, essentially establishing a mathematical representation of LLM throughput behavior.
  \dlist{%
  Parameter extrapolation with \ml} 
  ALA leverages \ml to predict the analytical model's parameters for unobserved workload configurations. This enables accurate extrapolation from the limited initial observed data.
  \dlist{%
  Search space exploration for error analysis} ALA employs simulated annealing to systematically explore the search space of different training point combinations. During this process, ALA logs the relationship between training point selection and resulting prediction errors, creating a map of prediction accuracy across different subspaces.
  \dlist{%
  Error prediction model development} Using the simulated annealing logs, ALA then builds a \ml based error predictor trained on these subspace observations. The resulting error predictor estimates inference performance prediction accuracy.
  \dlist{%
  Uncertainty quantification} ALA implements an uncertainty quantification system that calculates a confidence measure based on vector space similarity between new workloads and prior observations. This mechanism provides a reliability assessment for ALA's predictions.
  \dlist{%
  Performance characteristic prediction} Finally, ALA uses unobserved workloads, for which statistical performance characteristics are not initially known, together with the previously built error model on the simulated annealing logs to predict the analytical model’s characteristics and in turn predict the LLM inference statistical performance characteristics, including the confidence measure.

ALA focuses on predicting throughput, a key statistical performance characteristics most relevant to LLM inference systems, to enable developers to optimize deployment configurations without benchmarking every possible workload scenario. Its comprehensive integration of analytical modeling with \ml techniques provides a powerful methodology for predicting LLM performance across the wide spectrum of possible deployment configurations, significantly reducing the need for exhaustive benchmarking while maintaining reliable performance estimates.

\subsection{Generalized Exponential Model for Throughput Prediction}

\begin{figure}[t]
    \centering
        \includegraphics[scale=0.4,trim=0cm 0.33cm 0cm 0.95cm,clip]{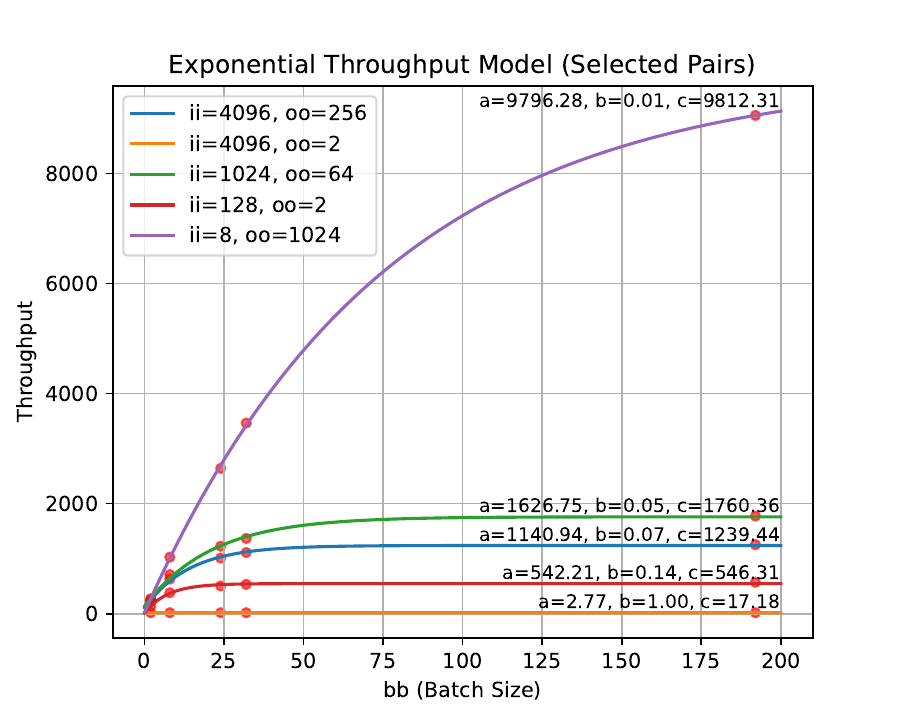}
    \caption{Exponential Models for Throughput in LLAMA} 
    \label{fig:expmodel}
\end{figure}

\noindent
In many LLM inference workloads, as we determined by examining the experimental results discussed in Section~\ref{sec:experiments}, system throughput exhibits an asymptotic growth pattern with respect to batch size. Initially, increasing the batch size leads to significant performance gains; however, due to hardware constraints, memory bandwidth limitations, and parallelization inefficiencies, the throughput eventually saturates as shown in Fig.~\ref{fig:expmodel}. To model this behavior effectively, we employ a \textit{Generalized Exponential Model} in \textbf{Algorithm \ref{alg:exp_model}}.

\begin{algorithm}
    \caption{Generalized Exponential Model for Throughput ($thpt$) as function of batch size ($bb$)}
    \label{alg:exp_model}
    \KwIn{$bb$: Input feature, $a, b, c$: Model parameters}
    \KwOut{$thpt$: Predicted throughput}

    \tcp{Compute throughput using the exponential model}
    $thpt \gets c - a \cdot e^{-b \cdot bb}$\;

    \Return{$thpt$}\;
\end{algorithm}

\noindent
The throughput $thpt$ is computed using the following exponential equation: 
%
%
$thpt = c - a \cdot e^{-b \cdot bb}$, 
where:
\begin{itemize}
    \item $bb$ represents the \textit{batch size}, an input feature.
    \item $a$ is a \textit{scaling parameter} that determines the magnitude of the initial throughput improvement.
    \item $b$ is the \textit{rate parameter}, controlling how quickly the throughput approaches its asymptotic limit.
    \item $c$ represents the \textit{saturation throughput}, denoting the upper bound on throughput as the batch size increases.
\end{itemize}

\noindent

\begin{algorithm}
    \caption{Building Exponential Database}
    \label{alg:exp_db}
    \KwIn{$\mathcal{D}$: Benchmark subdataset with  $(ii, oo, bb, thpt)$}
    \KwOut{$\mathcal{P}$: Parameter database, $\mathcal{T}$: Training parameters}

    $\mathcal{P} \gets \emptyset$, $\mathcal{T} \gets \emptyset$\;

    \ForEach{$(ii, oo) \in$ unique pairs in $\mathcal{D}$}{
        $G \gets$ subset of $\mathcal{D}$ where $(ii, oo)$ matches\;
        $bb \gets$ values of $bb$ in $G$\;
        $thpt \gets$ values of $thpt$ in $G$\;

        \If{$|bb| > 1$}{
            $thpt_{p10}, thpt_{p90} \gets 10^{th}, 90^{th}$ percentile of $thpt$\;
            $b_{p10}, b_{p90} \gets 10^{th}, 90^{th}$ percentile of $bb$\;
            $b_{p90} \gets \max(b_{p90}, b_{p10} + \epsilon), \quad \epsilon = 10^{-3}$\;

            \tcp{Initial parameter estimation}
            $a_0 \gets \max(thpt_{p90} - thpt_{p10}, 10^{-5})$\;
            $b_0 \gets 1 / \max(b_{p90} - b_{p10}, 10^{-5})$\;
            $c_0 \gets \max(thpt_{p90}, 10^{-5})$\;
        }
        \Else{
            $a_0 \gets 1.0, \quad b_0 \gets 0.001, \quad c_0 \gets 0.0$\;
        }

        \tcp{Curve fitting using nonlinear optimization}
        $\hat{a}, \hat{b}, \hat{c} \gets$ Optimize $thpt = c - a e^{-b \cdot bb}$ \; 

        \If{Optimization successful}{
            $\mathcal{P}[(ii, oo)] \gets (\hat{a}, \hat{b}, \hat{c})$\;
            Append $(ii, oo, \hat{a}, \hat{b}, \hat{c})$ to $\mathcal{T}$\;
        }
    }

    \Return{$\mathcal{P}, \mathcal{T}$ if $\mathcal{T}$ is not empty, else $\emptyset$}\;
\end{algorithm}

\subsection{Building the Exponential Parameter Database}

\noindent
Accurately modeling throughput in LLM inference workloads requires capturing the non-linear relationship as depicted in Fig.~\ref{fig:expmodel}. We construct an \textit{Exponential Parameter Database}, where workload-specific throughput trends are represented using the generalized exponential function. The database is built by fitting the model parameters $(a, b, c)$ for each unique hardware-software parameter configuration, as we describe in Section \ref{sec:combotraining}.
As detailed in \textbf{Algorithm \ref{alg:exp_db}}, given a benchmarked dataset $\mathcal{D}$ with features $(ii, oo, bb, thpt)$—where $ii$ represents input instance size, $oo$ denotes output instance size, $bb$ corresponds to batch size, and $thpt$ is the measured throughput, we first partition the dataset based on unique $(ii, oo)$ pairs, grouping all throughput measurements for each workload configuration. For each group, the algorithm estimates parameters using the 10th and 90th percentiles: $a_0$ is set as the difference between these percentiles for throughput, $b_0$ as the inverse of their difference for batch size, and $c_0$ as the maximum observed 90th percentile throughput. The algorithm leverages the 10th and 90th percentiles to estimate initial parameters because they provide a robust measure of variability while mitigating the influence of outliers. The difference between these percentiles for throughput ($a_0$) captures the spread of observed performance, offering a reasonable approximation of how throughput changes across batch sizes. The inverse difference for batch size ($b_0$) ensures numerical stability, preventing extreme values that could skew model fitting. The 90th percentile throughput ($c_0$) is used as an upper bound since it represents a realistic high-performance estimate under favorable conditions. If only a single batch size exists, default values $(a_0 = 1.0, b_0 = 0.001, c_0 = 0.0)$ are assigned. Next, the generalized exponential model, $thpt = c - a e^{-b \cdot bb}$, is fitted using nonlinear least squares optimization, ensuring numerical stability with bounded constraints. The fitted parameters are stored in $\mathcal{P}$, and the training parameters are appended to $\mathcal{T}$. Finally, the function returns $\mathcal{P}$ and $\mathcal{T}$, or an empty set if no valid estimates were obtained.
The exponential database constructed using this method is depicted in Fig.~\ref{fig:expmodel}, serves as a foundation for the \ml augmentation  that we explain next.

\begin{algorithm}
    \caption{Parameter Prediction Model Training}
    \label{alg:xgboost_train}
    \KwIn{$\mathcal{T}$: Training dataset with parameters $(ii, oo, a, b, c)$}
    \KwOut{$\mathcal{M}$: Trained XGBoost model}

    \lIf{$\mathcal{T} = \emptyset$}{\Return{None}}

    \tcp{Feature Engineering}
    \ForEach{$(ii, oo) \in \mathcal{T}$}{
        $\log_{ii} \gets \log(1 + ii)$\;
        $\log_{oo} \gets \log(1 + oo)$\;
        $\log_{bb} \gets \log(1 + ii / oo)$\;
        $ii\_oo\_ratio \gets ii / (oo + 1)$\;
        $ii\_ii\_ratio \gets ii / (ii + 1)$\;
    }

    \tcp{Prepare Features and Targets}
    $X \gets [ii, oo, \log_{ii}, \log_{oo}, \log_{bb}, ii\_oo\_ratio, ii\_ii\_ratio]$\;
    $y \gets [a, b, c]$\;


    \tcp{Train Multi-Output XGBoost Model}
    $\mathcal{M} \gets$ MultiOutputRegressor(XGBoost)\;
    Fit $\mathcal{M}$ using $(X, y)$\;

    \Return{$\mathcal{M}$}\;
\end{algorithm}

\subsection{Augmenting Learning to Predict Throughput Models}  \label{sec:combotraining}

\begin{figure}
    \centering
        \includegraphics[scale=0.4,trim=0 0.3cm 0 0.9cm,clip]{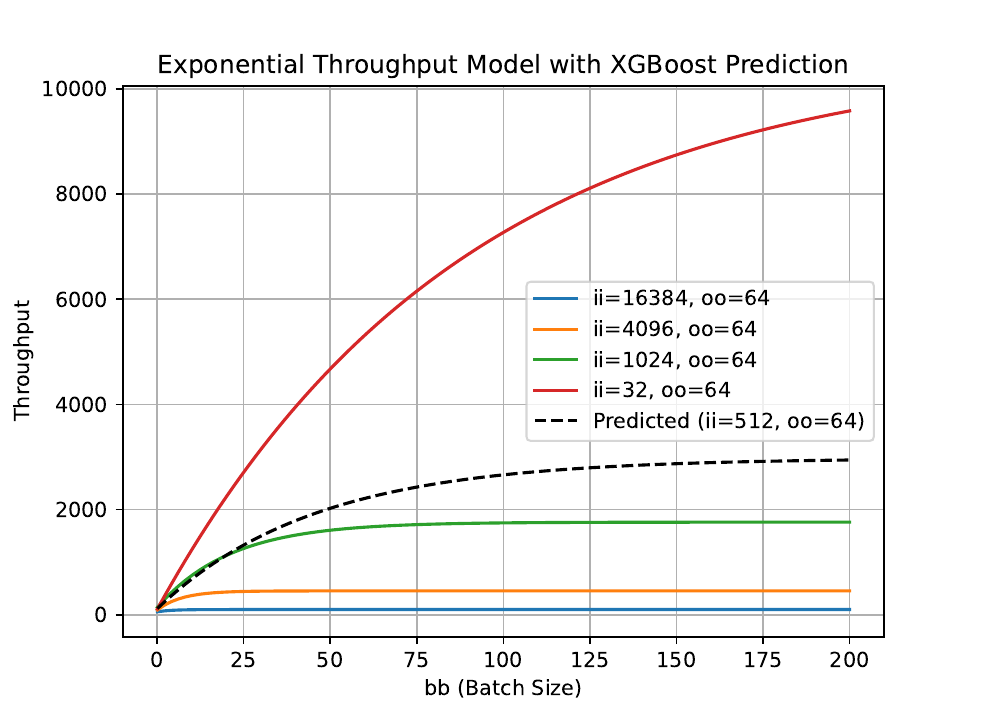}
    \caption{Exponential Models for Throughput in LLAMA} 
    \label{fig:expmodelpred}
\end{figure}

\noindent
For previously unseen $(ii, oo)$ pairs, we utilize an XGBoost-based multi-output regression model to generalize the relationship between workload features and the exponential model parameters $(a, b, c)$. Our methodology is generic and other \ml models can be uses as well.
\textbf{Algorithm \ref{alg:xgboost_train}} describes the training procedure for an XGBoost based model, which takes as input the parameter dataset $\mathcal{T}$ consisting of $(ii, oo, a, b, c)$ tuples. Next, a feature engineering step is performed to enhance the model's expressiveness including computing logarithmic transformations of input and output sizes to capture scale variations, deriving a batch scaling feature to reflect workload characteristics, and computing input-output and normalized input ratios to mitigate extreme values. The input-output ratio ($\mathit{ii\_oo\_ratio}$) provides insight into the relationship between input and output sizes, helping the model generalize across different workload distributions. Similarly, the normalized input ratio ($\mathit{ii\_ii\_ratio}$) is a nonlinear transformation that normalizes input size into a stable and interpretable range, capturing diminishing effects in model behavior with respect to increasing input size.
Three separate XGBoost regressors are trained to predict $(a, b, c)$, resulting in a model $\mathcal{M}$ that generalizes to unseen configurations, improving throughput prediction while preserving interpretability. An example of the prediction of this algorithm is depicted in Figure \ref{fig:expmodelpred}.

\begin{algorithm}
    \caption{Training  for Parameter Combinations}
    \label{alg:train_save_models}
    \KwIn{$\mathcal{D}$: Benchmark Data 
    }
    \KwOut{Trained XGBoost models and parameter databases}

    \tcp{Extract Unique Parameter Combinations}
    $\mathcal{C} \gets$ Unique combinations of $(acc, acc\_count, back, cpu, cpu\_count, dev, mode, model, prec)$ in $\mathcal{D}$\;

    \ForEach{$\mathbf{c} \in \mathcal{C}$}{
        \tcp{Filter dataset based on parameter combination}
         $\mathcal{D_C} \gets$ Subset of $\mathcal{D}$ where attributes match $\mathbf{c}$\;




        \tcp{Train Model}
        $\mathcal{H} \gets$ Select columns $(bb, ii, oo, thpt)$ from $\mathcal{D}_\mathcal{C}$\;
        $(\mathcal{P}, \mathcal{T_P}) \gets$ \textbf{build\_exponential\_database}$(\mathcal{H})$\;
        $\mathcal{M} \gets$ \textbf{train\_xgboost}$(\mathcal{T_P})$\;



    }
\end{algorithm}


\noindent
\textbf{Algorithm \ref{alg:train_save_models}} outlines the process of training and storing XGBoost models for different parameter configurations in LLM inference workloads. It first filters the benchmark dataset $\mathcal{D}$ to extract unique parameter combinations (example demonstrated in Algorithm \ref{alg:train_save_models}, can include any other combination or may not include all listed, and hence is generic in nature), forming the set $\mathcal{C}$. For each combination, it selects a subset $\mathcal{D_C}$, and collects training data $\mathcal{H}$. The \textbf{build\_exponential\_database} function estimates throughput model parameters, which are then used to train an XGBoost model via \textbf{train\_xgboost}. 

\begin{algorithm}
    \caption{Prediction with Parameter Lookup}
    \label{alg:xgb_prediction}
    \KwIn{$bb$: Batch size, $ii$: Input size, $oo$: Output size}
    \KwOut{$\hat{T}$: Predicted throughput}

    \tcp{Load Pre-trained Model and Parameter Database}
    Load XGBoost model $\mathcal{M}$\;
    Load parameter database $\mathcal{P}$\;

    \tcp{Check if $(ii, oo)$ exists in parameter database}
    \If{$(ii, oo) \in \mathcal{P}$}{
        Retrieve $(\hat{a}, \hat{b}, \hat{c})$ from $\mathcal{P}$\;
    }
    \Else{
        \tcp{Predict parameters using XGBoost model}
        $(\hat{a}, \hat{b}, \hat{c}) \gets \textbf{predict\_params}([bb, ii, oo], \mathcal{M})$\;
    }

    \tcp{Compute Throughput using the Exponential Model}
    $\hat{T} \gets \hat{c} - \hat{a} \cdot e^{-\hat{b} \cdot bb}$\;


    \Return $\hat{T}$\;
\end{algorithm}

\subsection{Throughput Prediction}

\noindent
\textbf{Algorithm \ref{alg:xgb_prediction}}  predicts throughput by first checking if the test input, consisting of batch size ($bb$), input size ($ii$), and output size ($oo$), exists in the parameter database $\mathcal{P}$. If a matching $(ii, oo)$ entry exists, it uses the corresponding precomputed exponential model parameters to estimate throughput directly. Otherwise, the XGBoost model predicts the exponential model parameters, which are then used to compute the estimated throughput $\hat{T}$. 
In the subsequent sections, we leverage the above algorithms coupled with Simulated Annealing to weave through various subspaces of the training dataset and generalize an error predictor towards other unknown workloads.

\begin{algorithm}
    \caption{Simulated Annealing for State Space Exploration}
    \label{alg:simulated_annealing}
    \KwIn{$\mathcal{D}$: Benchmark Dataset, $\mathcal{S}_0$: Initial subsets, $\tau$: Initial temperature, $\alpha$: Cooling rate, $N$: Max iterations}
    \KwOut{$\mathcal{L}$ : Iteration Logs}

    $\mathcal{S}^* \gets \mathcal{S}_0$, $\mathcal{E}^* \gets \textbf{EvaluateSubset}(\mathcal{S}^*)$\;
    
    \For{$i \gets 1$ to $N$}{
        $\tau \gets \alpha \cdot \tau$\; 
        
        \tcp{Generate a new candidate subset}
        $\mathcal{S}' \gets \textbf{ModifySubset}(\mathcal{S}^*)$\;
        
        \tcp{Evaluate the new subset}
        $\mathcal{E}' \gets \textbf{EvaluateSubset}(\mathcal{S}')$\;
        
        \tcp{Acceptance criterion}
        \If{$\mathcal{E}' < \mathcal{E}^*$ \textbf{or} $\exp\left(\frac{\mathcal{E}^* - \mathcal{E}'}{\tau}\right) > \textbf{rand}()$}{
            $\mathcal{S}^* \gets \mathcal{S}'$, $\mathcal{E}^* \gets \mathcal{E}'$\;
        }
        
        \tcp{Store iteration data}
        Append $(\mathcal{S}', \mathcal{E}')$ to $\mathcal{L}$\;
        
    }

    \Return $\mathcal{L}$\;
\end{algorithm}

\subsection{Simulated Annealing for Optimal Subset Selection}

\noindent
\textbf{Algorithm \ref{alg:simulated_annealing}} implements a simulated annealing approach to explore the subset training sets in conjunction with the earlier algorithms that build generalized exponential throughput models and the learning based parameter predictor. We use median percentage error for this purpose, however, our algorithm is general and any error measure can be leveraged.  The algorithm initializes with a starting subset $\mathcal{S}_0$ and computes its initial error, $\mathcal{E}^*$. The annealing process iterates up to a predefined maximum number of steps, $N$, gradually reducing the temperature $\tau$ using a cooling rate $\alpha$.
At each iteration, it generates a new candidate subset $\mathcal{S}'$ by modifying the current best subset $\mathcal{S}^*$ by randomly adding or deleting elements from the $(ii, oo, bb)$ training data followed by evaluating its corresponding error, $\mathcal{E}'$. The new subset is accepted if it achieves a lower error than $\mathcal{E}^*$ or, with a probability given by $\exp\left(\frac{\mathcal{E}^* - \mathcal{E}'}{\tau}\right)$, allowing occasional acceptance of worse solutions to enhance search space exploration.
Each iteration logs the evaluated subset and its error to track the search process. The algorithm terminates when the maximum iteration count is reached. The algorithm thus examines different training data subspaces and their associated errors. We leverage these logs to first train an error predictor based on the assumption that the workload follows a similar generalized exponential throughput model and then relax the assumption by adding a generalizability adaptation framework based on the vector space similarity of the workloads.

\begin{algorithm}
    \caption{Error Predictor Training}
    \label{alg:mape_predictor}
    \KwIn{$\mathcal{L}$: Simulated annealing iteration history}
    \KwOut{$\mathcal{F}$: Trained error predictor}

    \tcp{Initialize datasets}
    $\mathcal{X} \gets \emptyset$, $\mathcal{Y} \gets \emptyset$\;

    \tcp{Encode each subset and collect corresponding MAPE}
    \ForEach{$(\mathcal{S}, \mathcal{M}) \in \mathcal{L}$}{
        $\mathcal{S}_{ii}, \mathcal{S}_{bb}, \mathcal{S}_{oo} \gets \mathcal{S}$\;
        $\textbf{x} \gets \emptyset$\;
        
        \ForEach{$v \in \texttt{unique\_ii}$}{
            Append $1$ to $\textbf{x}$ if $v \in \mathcal{S}_{ii}$, else append $0$\;
        }
        \ForEach{$v \in \texttt{unique\_bb}$}{
            Append $1$ to $\textbf{x}$ if $v \in \mathcal{S}_{bb}$, else append $0$\;
        }
        \ForEach{$v \in \texttt{unique\_oo}$}{
            Append $1$ to $\textbf{x}$ if $v \in \mathcal{S}_{oo}$, else append $0$\;
        }

        Append $\textbf{x}$ to $\mathcal{X}$\;
        Append $\mathcal{M}$ to $\mathcal{Y}$\;
    }

    \tcp{Train XGBoost regression model}
    $\mathcal{F} \gets \textbf{TrainXGBoost}(\mathcal{X}, \mathcal{Y})$\;

    \Return $\mathcal{F}$\;
\end{algorithm}

\subsection{Error Predictor Training and Evaluation}

\noindent
\textbf{Algorithm \ref{alg:mape_predictor}} describes the process of training an XGBoost-based predictor for estimating the error (e.g., MAPE) of new workload subsets. The predictor is trained using the iteration history $\mathcal{I}$ collected during the simulated annealing process. For each evaluated subset $\mathcal{S}$, the training data is constructed by encoding the subset into a binary feature vector. This encoding is performed by checking the presence of each element from the universal sets of input (\texttt{unique\_ii}), bottleneck (\texttt{unique\_bb}), and output (\texttt{unique\_oo}) workloads. If an element is included in $\mathcal{S}$, a 1 is appended to the feature vector; otherwise, a 0 is added. The corresponding observed error value $\mathcal{M}$ is recorded as the target.
Once the full dataset of encoded features $\mathcal{X}$ and target values $\mathcal{Y}$ is assembled, an XGBoost regression model is trained using the \textbf{TrainXGBoost} function. The resulting model $\mathcal{F}$ is returned and can be used to predict the error of previously unseen workload configurations, enabling more informed subset selection.

\begin{algorithm}
    \caption{Predicting Error and Confidence Estimation}
    \label{alg:predict_mape_unknown}
    \KwIn{$\mathcal{D}_{\text{new}}$: Unknown dataset, $\mathcal{F}$: Trained MAPE predictor, $\mathcal{L}$: Simulated annealing iteration history}
    \KwOut{$\hat{\mathcal{M}}$: Predicted error, $c$: Confidence score}

    \tcp{Extract unique subset signature from new dataset}
    $\mathcal{S}_{\text{new}} \gets \textbf{ExtractUniqueParameterCombinationSets}(\mathcal{D}_{\text{new}})$\;

    \tcp{Encode the new dataset subset, vector of encodings}
    $\mathcal{X}_{\text{new}} \gets \textbf{EncodeCombinationsSets}(\mathcal{S}_{\text{new}})$\;

    \tcp{Predict Error using the trained model, vector of errors returned}
    $\hat{\mathcal{M}} \gets \mathcal{F}(\mathcal{X}_{\text{new}})$\;

    \tcp{Compute distance to nearest known training subset for each combination}
    $\mathcal{V} \gets [\textbf{EncodeSubset}(\mathcal{S}) \mid (\mathcal{S}, \mathcal{M}) \in \mathcal{L}]$\;
    $d_{\text{min}} \gets \min\limits_{\mathcal{V}_i \in \mathcal{V}} \textbf{Distance}(\mathcal{X}_{\text{new}}, \mathcal{V}_i)$\;

    \tcp{Compute confidence score based on distance}
    $c \gets \frac{1}{1 + d_{\text{min}}}$\;

    \Return $(\hat{\mathcal{M}}, c)$\;
\end{algorithm}

\textbf{Algorithm \ref{alg:predict_mape_unknown}} estimates the error for an unseen LLM workload using the error predictor and quantifies prediction confidence based on similarity to previously evaluated subsets. The algorithm first extracts unique parameter combinations from the new dataset and encodes them into feature representations as discussed in Algorithm \ref{alg:mape_predictor}. It then uses the predictor to estimate the error. To assess confidence in this prediction, the algorithm computes the minimum distance between the encoded new workload and all previously encountered subsets from the iteration history. This distance is the vector space distance between the $(ii, oo, bb, thpt)$ features of the two sets. Since the two sets are of unequal size, we abstract this setting by comparing the distance between the histograms of the two vector spaces. We compute the histogram for each feature, i.e., $ii$ , $bb$, .. and then compute the distance measure. We use the cosine similarity distance for this purpose and the final distance is the average distance for each feature. The final output includes the predicted error along with a confidence score computed as $c \gets \frac{1}{1 + d_{\text{min}}}$, which inversely scales with distance — smaller distances yield higher confidence. This formulation allows the model to quantify uncertainty in extrapolating to unfamiliar LLM configurations.

\section{Evaluation and Discussion}
\label{sec:experiments}

\begin{figure}
    \centering
    \begin{subfigure}{0.3\columnwidth}
        \includegraphics[width=\textwidth]{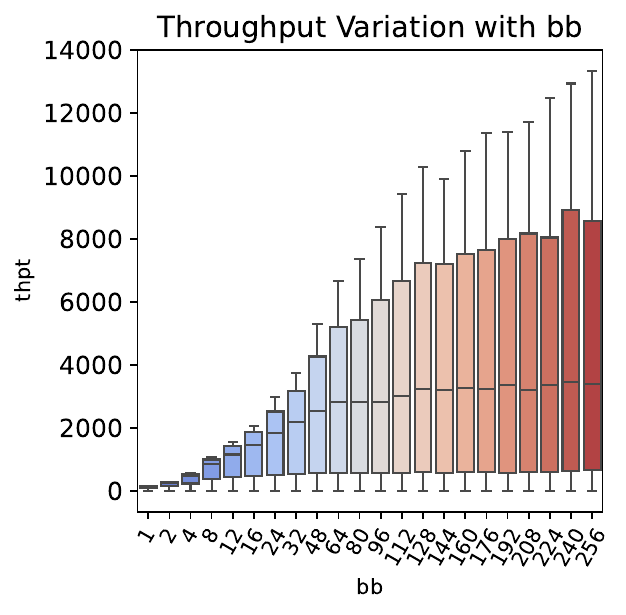}
        \caption{Throughput vs Batch Size}
        \label{fig:thpt_vs_bb}
    \end{subfigure}
    \hfill
    \begin{subfigure}{0.3\columnwidth}
        \includegraphics[width=\textwidth]{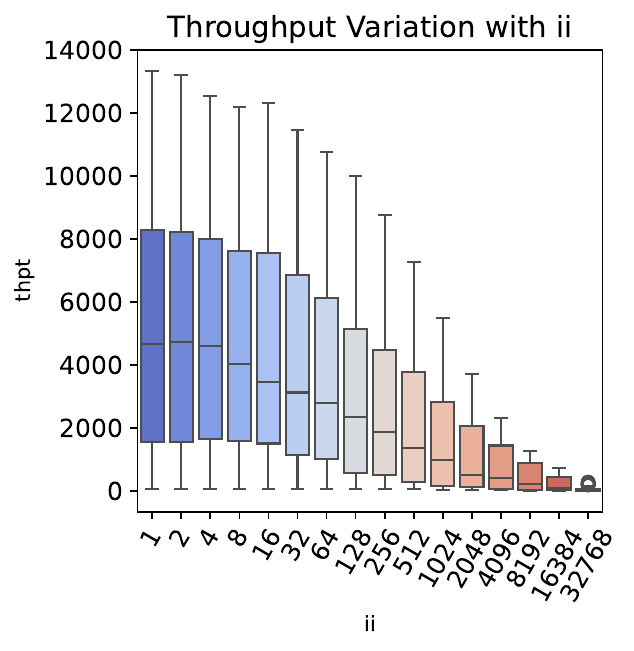}
        \caption{Throughput vs Input Token Size}
        \label{fig:thpt_vs_ii}
    \end{subfigure}
    \hfill
    \begin{subfigure}{0.3\columnwidth}
        \includegraphics[width=\textwidth]{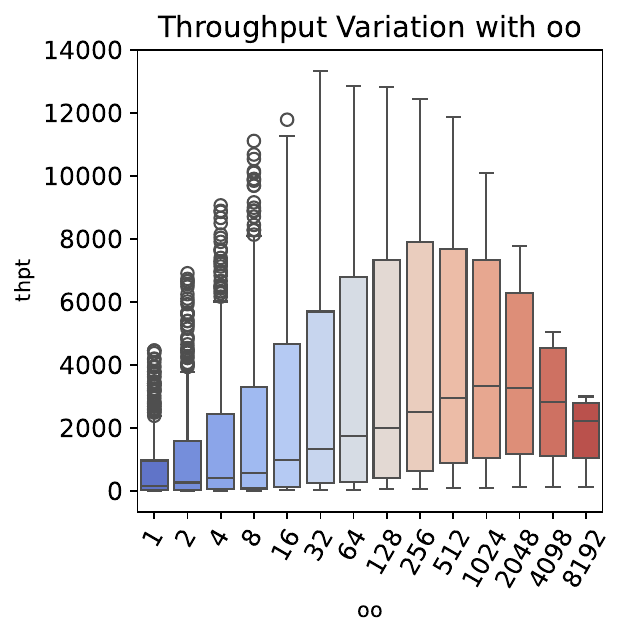}
        \caption{Throughput vs Output Token Size}
        \label{fig:thpt_vs_oo}
    \end{subfigure}
    \caption{Dataset Statistical Characteristics showing throughput variation with (a) batch size, (b) input token size, and (c) output token size.}
    \label{fig:datasetstats}
\end{figure}


\subsection{Experimental Setup}
\noindent
We conduct our experiments using two datasets, an in-house dataset we generated and an open source dataset, LLM-inference-bench dataset \cite{chitty2024llm}. To build our in-house dataset, we used a custom inference benchmarking framework that evaluates model performance across various configurations. Key parameters—such as model architecture, precision, sequence length, output size (tokens), and batch size—are systematically varied. Each combination is tested 5–10 times to account for runtime variability. We use throughput (tokens/sec) as the primary metric to assess LLM inference scalability. We primarily use LLaMA 3.1–8B, served with VLLM on NVIDIA H100 GPUs. The explored parameter space for input size, output size, and batch size is summarized in Fig.~\ref{fig:datasetstats}. To enable large-scale performance analysis, our in-house dataset comprises approximately 4,800 data points by systematically varying input token size, output token size, and batch size—including configurations to large values that has been relatively less explored. To the best of our knowledge, this is one of the first contributions to characterize inference performance at such scale, filling a critical gap in existing benchmarking datasets. The LLM-inference-bench dataset includes several LLM inference frameworks and models from LLaMA, Mistral, and Qwen families with 7B and 70B parameters with batch sizes ranging from 1 to 64, and input / output length from 128 to 2048. Fig.~\ref{fig:datasetstats} illustrates the statistical properties of the collected datasets. Specifically, Figs.~\ref{fig:datasetstats} (a), (b),  and (c) depict the throughput variability across different operational settings. The datasets exhibit significant variations as batch size increases  non-linearly, suggesting heterogeneity in service behavior and highlighting the necessity for robust modeling strategies.



\subsection{Results}

\begin{figure}
    \centering
    \begin{subfigure}{0.48\columnwidth}
        \centering
        \includegraphics[scale=0.55,trim=0.6cm 8.5cm 8.0cm 1.2cm,clip]{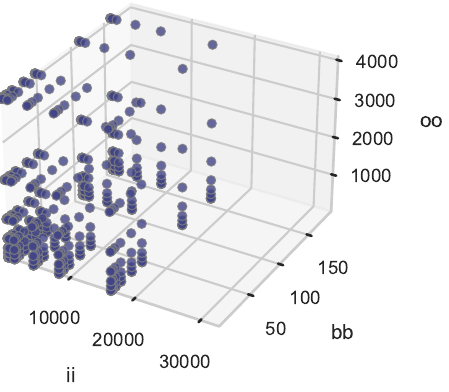}
        \caption{Experiment 1}
    \end{subfigure}
    \hfill
    \begin{subfigure}{0.44\columnwidth}
        \centering
        \includegraphics[scale=0.55,trim=9.2cm 8.5cm 0cm 1.2cm,clip]{Res/statisticstrainset.pdf}
        \caption{Experiment 2}
    \end{subfigure}
    \\[0.5em]
    \begin{subfigure}{0.48\columnwidth}
        \centering
        \includegraphics[scale=0.55,trim=0.6cm 0cm 8.0cm 9.8cm,clip]{Res/statisticstrainset.pdf}
        \caption{Experiment 3}
    \end{subfigure}
    \hfill
    \begin{subfigure}{0.44\columnwidth}
        \centering
        \includegraphics[scale=0.55,trim=9.2cm 0cm 0cm 9.8cm,clip]{Res/statisticstrainset.pdf}
        \caption{Experiment 4}
    \end{subfigure}
    \caption{Distribution of Training Sets} 
    \label{fig:distributiontrainset}
\end{figure}

\begin{figure}
    \centering
    \begin{subfigure}{0.45\columnwidth}
        \centering
        \includegraphics[width=\textwidth]{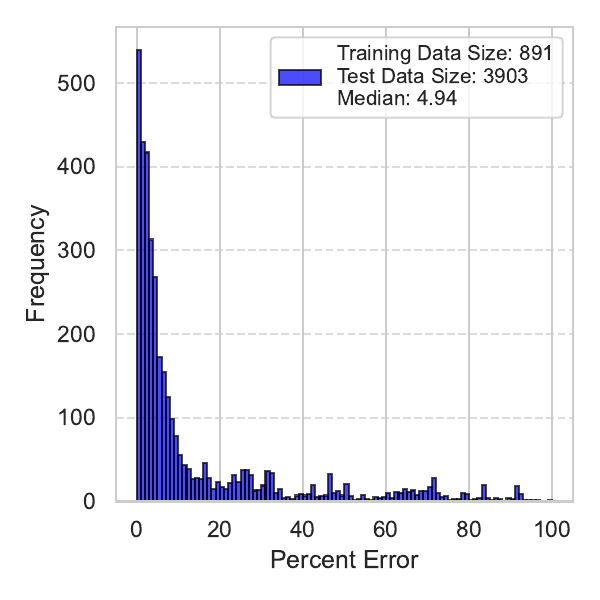}
        \caption{Experiment 1}
        \label{fig:hist1}
    \end{subfigure}
    \hfill
    \begin{subfigure}{0.45\columnwidth}
        \centering
        \includegraphics[width=\textwidth]{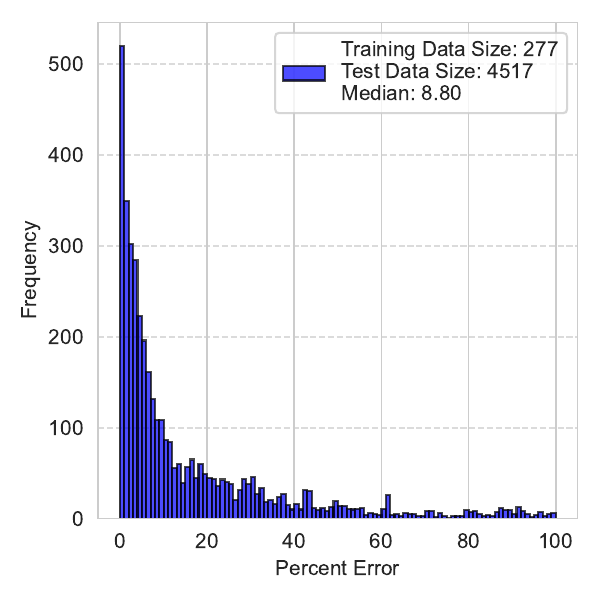}
        \caption{Experiment 2}
        \label{fig:hist2}
    \end{subfigure}
    \\[0.5em]
    \begin{subfigure}{0.45\columnwidth}
        \centering
        \includegraphics[width=\textwidth]{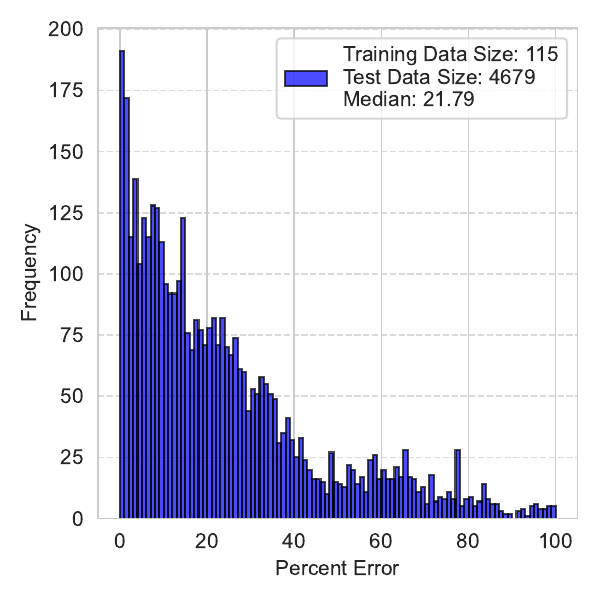}
        \caption{Experiment 3}
        \label{fig:hist3}
    \end{subfigure}
    \hfill
    \begin{subfigure}{0.45\columnwidth}
        \centering
        \includegraphics[width=\textwidth]{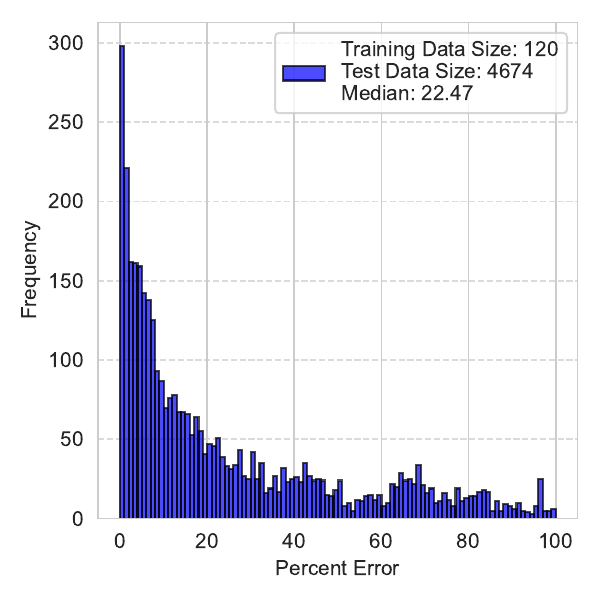}
        \caption{Experiment 4}
        \label{fig:hist4}
    \end{subfigure}
    \caption{Histogram of Errors for Different Training Sets}
    \label{fig:histograms}
\end{figure}

\noindent
To evaluate ALA, we answer the following research questions:

\qlist{RQ1: How does the training set impact the throughput prediction?} For this purpose, we define four experiments each with different training data covering different aspects of the state-space as depicted in Fig.~\ref{fig:distributiontrainset}. 
Fig.~\ref{fig:histograms} presents the distribution of prediction errors under the different training configurations. Across all training sets, the error distributions are right-skewed, with a concentration of low-error predictions but a noticeable long tail of higher errors. Notably, some configurations yield heavier tails, indicating that certain subsets may induce greater model instability or outlier behavior. Notably, Experiments 3 and 4 have the highest median error amongst the four sets. Experiment 3 does not include the training points from the larger batch sizes and hence the exponential nature of the relationship is not fully exposed leading to the larger errors. Similarly, Experiments 2 and 4 both include larger batch sizes, but differ in focus—Experiment 4 emphasizes sparsity across the overall range, while Experiment 2 concentrates on densely clustered metrics within specific regions. This density provides more consistent patterns for the model to learn from, reducing generalization error. As a result, the predictor can better infer the performance of similar configurations, leading to a lower error. Experiment 1 achieves the lowest error (4\%) due to its broad yet balanced coverage however highlighting the fact that large input sizes does not yield any significant advantage to the model.  The training set has a critical influence on the overall error percentage and thus the availability of LLM observability data can play a significant role in the overall prediction scope and accuracy.

%

\begin{figure}
    \centering
    \begin{subfigure}{0.45\columnwidth}
        \centering
        \includegraphics[width=\textwidth]{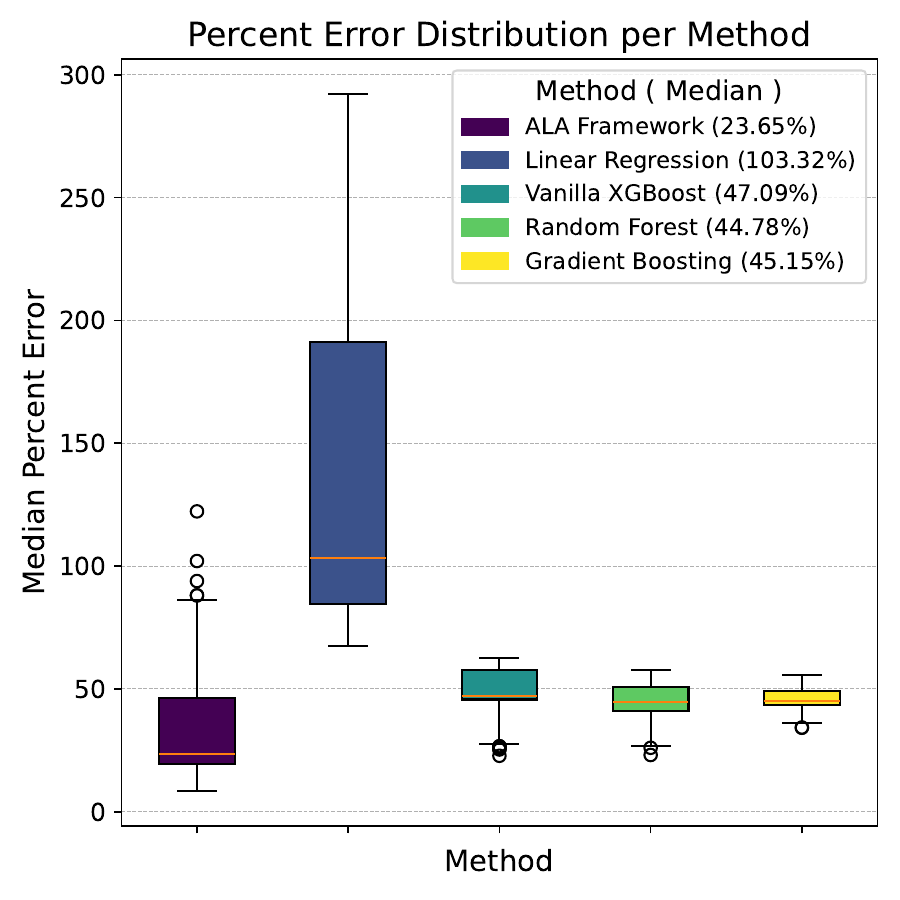}
        \caption{Percent Error Distribution}
        \label{fig:restat1}
    \end{subfigure}
    \hfill
    \begin{subfigure}{0.45\columnwidth}
        \centering
        \includegraphics[width=\textwidth]{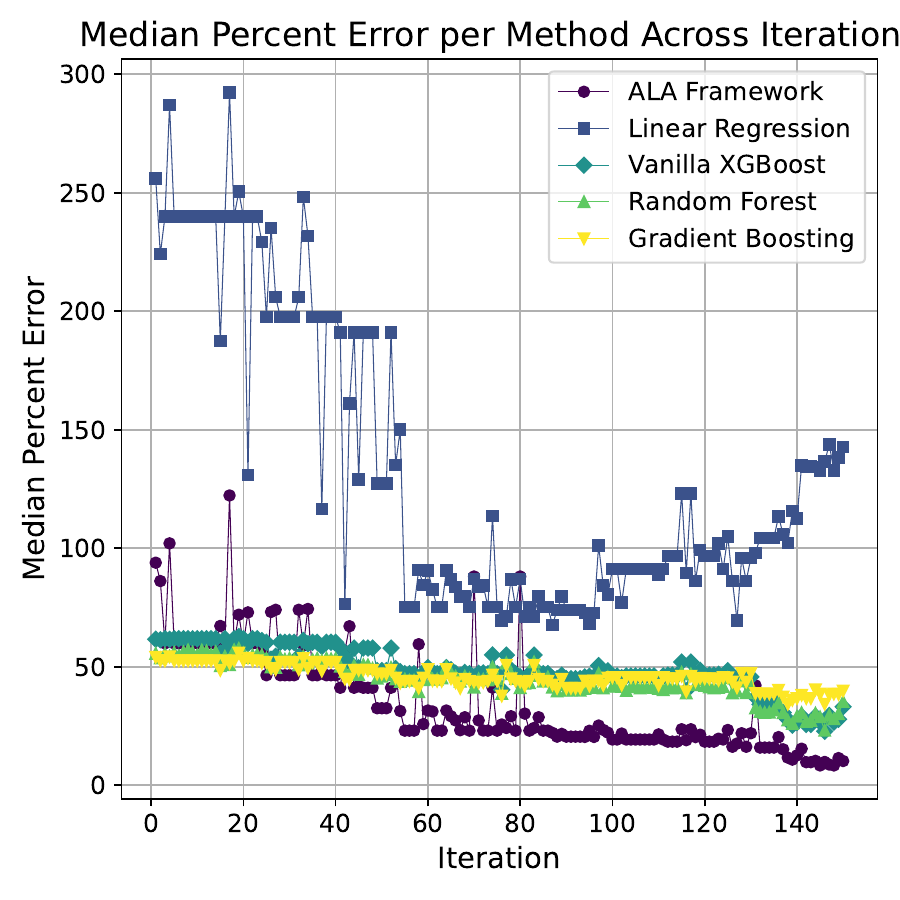}
        \caption{Simulated Annealing}
        \label{fig:restat2}
    \end{subfigure}
    \\[.5em]
    \begin{subfigure}{0.45\columnwidth}
        \centering
        \includegraphics[width=\textwidth]{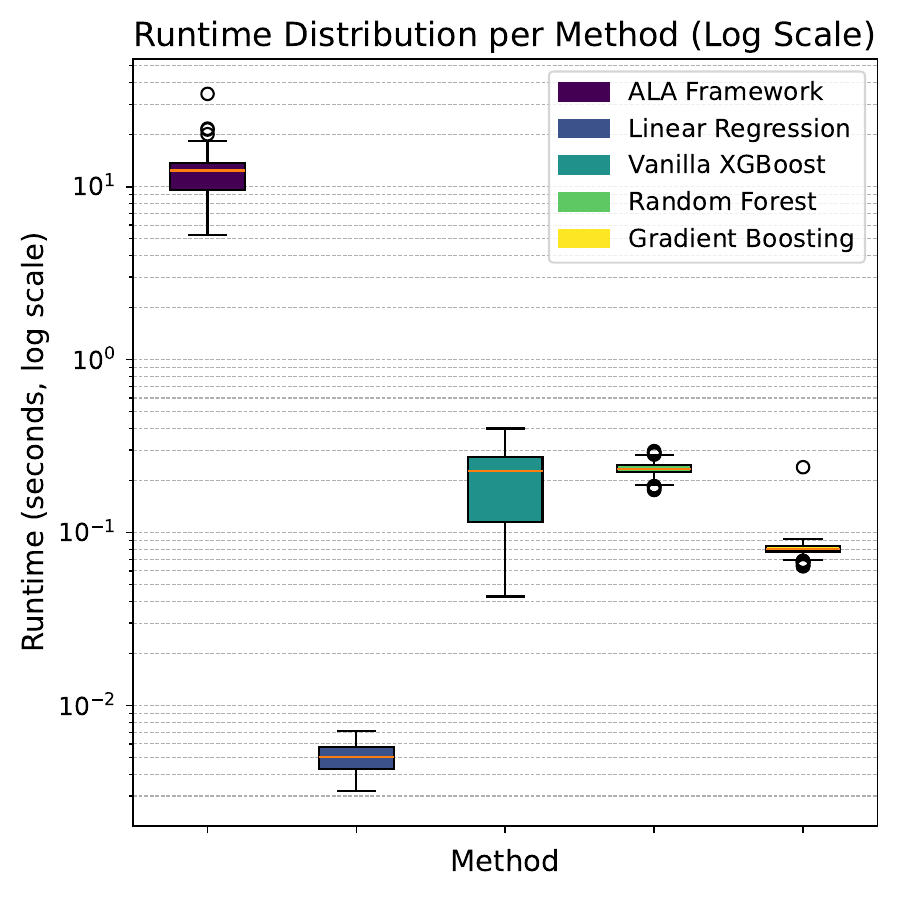}
        \caption{Runtime Variability}
        \label{fig:restat3}
    \end{subfigure}
    \hfill
    \begin{subfigure}{0.45\columnwidth}
        \centering
        \includegraphics[width=\textwidth]{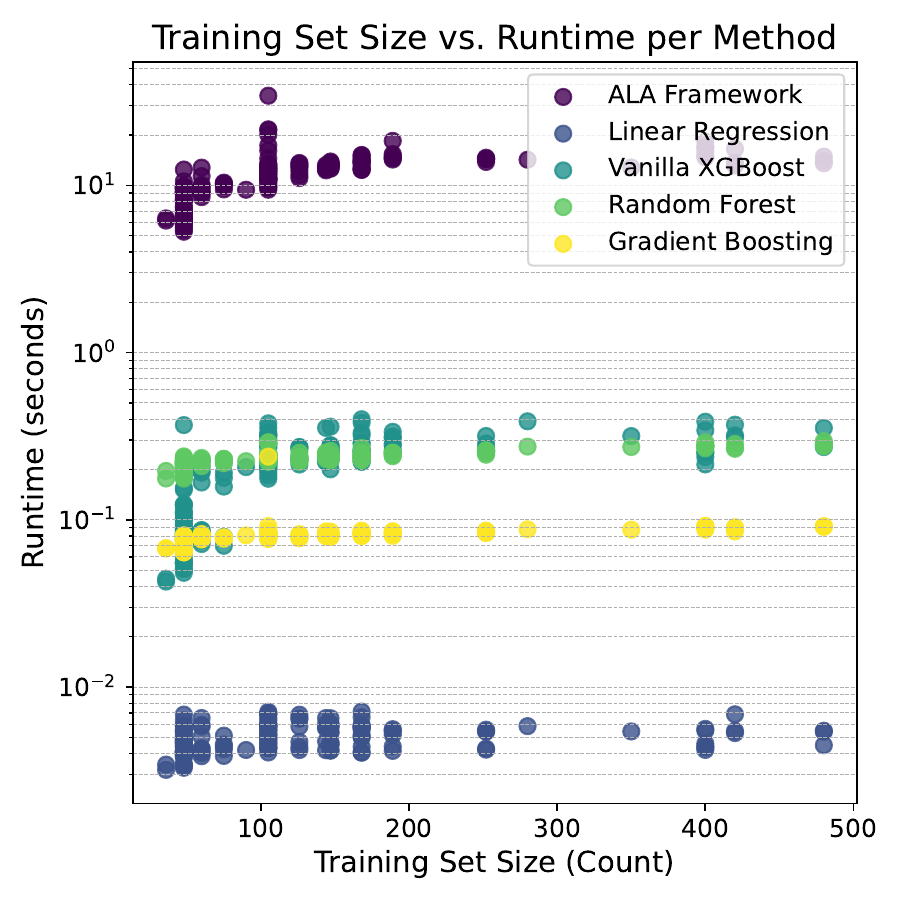}
        \caption{Impact of Training Set Size}
        \label{fig:restat4}
    \end{subfigure}
    \caption{Comparison with Baseline Approaches}
    \label{fig:restat}
\end{figure}

\qlist{RQ2: How does ALA fare in effectiveness and efficiency?}
Fig.~\ref{fig:restat} provides a comprehensive comparison between the ALA framework and several widely-used model baselines across multiple performance dimensions: (1) Linear Regression,  (2) Vanilla XGBoost: An implementation of Extreme Gradient Boosting without extensive hyperparameter optimization on the dataset itself as opposed to our analytical + XGB framework, (3) Random Forest and (4) Gradient Boosting. 
Unlike these baselines, the ALA framework introduces a multi-stage modeling pipeline tailored for complex and dynamic LLM inference environments. 
Fig.~\ref{fig:restat}(a) depicts the distribution of median percentage errors across methods reveals that ALA achieves the lowest median errors. This indicates not only improved central performance but also increased robustness across different experimental runs, a critical requirement for dependable LLM system modeling.
Fig.~\ref{fig:restat}(b) tracks the median percent prediction error over simulated annealing iterations. ALA exhibits a consistent and steady low error across iterations, demonstrating its effectiveness towards LLM modeling and prediction. In contrast, conventional baselines exhibit stagnation with high errors, highlighting their limited adaptability to the complex search space of LLM workloads. This figure also assesses how prediction errors evolve with varying training set sizes since in each iteration of the Simulated Annealing process, the training set is modified. ALA maintains lower prediction errors even when the available training data is reduced, demonstrating superior sample efficiency. Fig.~\ref{fig:restat}(c) shows that the proposed ALA framework has lower runtime variability than linear regression and is comparable to other baselines. However, ALA incurs longer training times due to its multi-stage process: fitting exponential models for each unique $(ii, oo)$ pair and then training an XGBoost model on their parameters. In contrast, baselines train faster due to simpler, dataset-wide modeling. Fig.~\ref{fig:restat}(d) indicates that training time across methods is largely unaffected by dataset size. For ALA, runtime depends more on the number of unique configurations than the total number of samples. Fig.~\ref{fig:restat}(a)-(b) reveal that the additional computational investment in ALA leads to significantly improved model fidelity, lower prediction error, and stronger generalization capabilities. 

\begin{figure}
    \centering
    \includegraphics[scale=0.36]{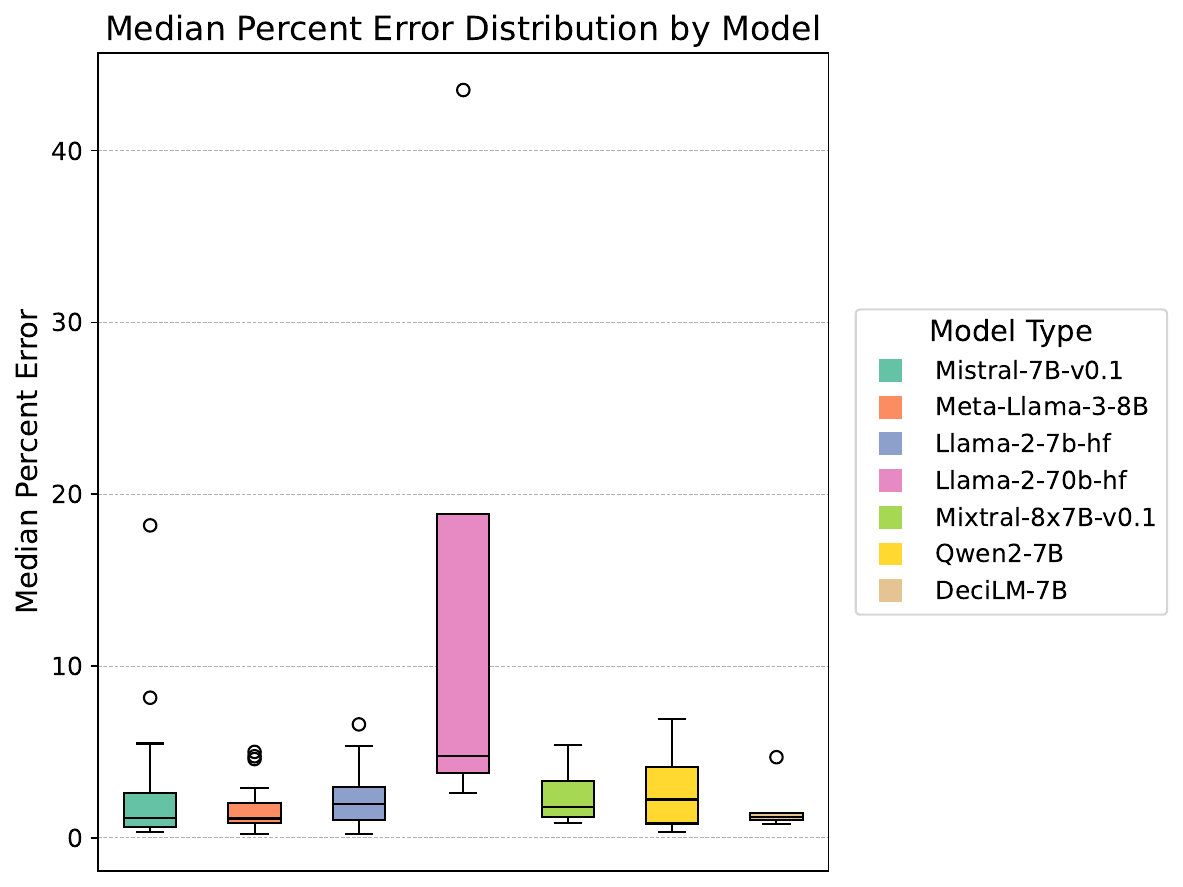}
    \caption{Error distribution observed in the ANL dataset.}
    \label{fig:anlplot}
\end{figure}

\qlist{RQ3: How does ALA fare with different LLM Models?}
Fig.~\ref{fig:anlplot} plots the median percentage error for the ANL dataset for the various model types with the ALA framework. For  each model type, we segregate the data into test and training and execute the ALA framework on that. The results demonstrate that ALA consistently achieves low median errors across all evaluated models. This generalizability across different models indicates that the exponential model effectively characterizes the throughput as a function of system resources and workload intensity across a variety of LLM models.
Notably, the LLama model exhibits a slightly larger spread in the error distribution compared to other models. This behavior can be attributed to the larger number of workload configurations available in the ANL dataset for LLama, which introduces greater inherent variability and a wider range of operational conditions. Despite this increased variability, the ALA framework maintains competitive median error levels, highlighting its ability to generalize even under more challenging and diverse data distributions.

\begin{table}
    \captionsetup{position=above}
    \caption{Performance prediction across different datasets with model confidence and actual error values.}\label{tab:predict}
    \centering\scriptsize
    \begin{tabularx}{\columnwidth}{@{}Xccc@{}}
        \toprule
        \textbf{Dataset} & \textbf{Predicted  Error} & \textbf{Confidence} & \textbf{Actual  Error} \\ \midrule
        LLAMA Subset & 15.34 & 0.99 & 15.03 \\ 
        Mistral 7B & 22.68 & 0.96 & 28.77 \\ 
        Qwen2-7B & 20.31 & 0.67 & 39.88 \\ \bottomrule
    \end{tabularx}
\end{table}


\qlist{RQ4: How well does ALA quantify uncertainty?}
Table~\ref{tab:predict} presents the performance prediction results of the XGBoost-based error predictor trained on logs generated via Simulated Annealing. Cosine similarity was employed as the metric to compare trace patterns during model training and inference. The predictor demonstrates high confidence and accurate error estimation on the LLAMA Subset and Mistral 7B datasets, which closely match the distribution of the training data. However, a significant discrepancy is observed for the Qwen2-7B dataset, where the predicted error underestimates the actual error. This is attributed to a hardware mismatch: Qwen2-7B was executed on an Intel PVC GPU, whereas the model was trained using execution traces collected from an NVIDIA H100 GPU. This highlights the the generalizability of the model towards the state space encompassed by the similarity metric exposed via the confidence measure.

\section{Related Work}
\label{sec:background}


\noindent
Previous studies (e.g., \cite{sarathi,agrawal2024vidur}) have observed the dramatic impact on LLM performance (both training and inference) of workload  characteristics (e.g., size of input and output tokens), the underlying model (e.g., LLM, precision), and backend configuration (e.g., GPU type, parallelism, batch size).  \cite{emani2023gpt2,llmonfrontier,YIN2021100005} studied leadership class super computers. \cite{emani2023gpt2,daydream,habitat,DLRM} focus on DNN and LLM training performance, which differs from 
inference predictions, since training jobs are (1) much longer than inference queries (run in the order of hundreds of milliseconds) and (2) have low workload variability, with fixed batch sizes and similar input and output sizes. Thus, inference predictions must provide much higher time granularity over a much broader configuration space.

DNN training simulation frameworks provide a cost effective way to test a broad configuration space without full benchmarking. However, profiling measurements are required to calibrate the simulations. \cite{habitat} extrapolates measurements taken on one GPU instance to multiple GPUs; \cite{daydream} uses profiling to identify bottlenecks and predict the affect of various optimizations on performance; \cite{proteus} simulates various parallelization strategies; and \cite{emani2023gpt2} applies transformer micro-benchmarks. Recently, Vidur \cite{agrawal2024vidur} provided a  simulation framework that accounts for the specific properties of LLM inference. It profiles three building block operators: token-level, sequence-level and communication and models various scheduling and parallelization schemes to simulate LLM inference on an entire workload trace. The simulator still requires profiling data for every input combination across all the operators. 

Since collecting actual measurements would be prohibitively expensive, Vidur relies on a runtime estimator that utilizes ML models to interpolate missing data points. With a similar goal, \cite{arise} predicts LLM inference performance utilizing ML models trained on historical workloads and benchmark metadata; it evaluates multiple regression models and applies hyper-parameter optimization. ML models work well for interpolation of missing data points but are less accurate for extrapolation, especially with new LLMs and/or new GPU models. Techniques like transfer learning and meta-learning are being explored to improve the generalization capabilities of ML models in LLM inference systems \cite{zhou2024survey}.

\cite{chitty2024llm} provide a benchmarking suite for LLM inference. Their empirical study further validates the relations between workload parameters ($bb,ii,oo$) and performance metrics (throughput and latency), as utilized in our ALA framework. 

Our work utilizes a hybrid approach that combines statistical modeling and ML prediction.
While analytical models provide insights into the relationships between various system parameters and statistical performance metrics, they often struggle to capture the complexity of real-world workloads. \ml-based approaches, on the other hand, can learn from historical data and make accurate predictions, but may lack interpretability and generalization capabilities. Our enhanced hybrid approach could provide inputs to simulators, such as Vidur, when predicting performance for entire inference trace.

\section{Conclusion and Future Work}
\label{sec:conclusion}

\noindent
In this work, we  introduced the Analytical with Learning Augmentation (ALA) framework, which combines analytical modeling with \ml to predict LLM inference performance for unseen workloads. Using simulated annealing for training data selection and \ml, ALA achieves accurate predictions with built-in confidence estimation via vector similarity. Experiments across varied workloads show low error and strong generalization. As part of future work, we aim to generalize this capability to support arbitrary subsets of the test space, enabling more flexible and targeted error estimation for diverse deployment scenarios.

\bibliographystyle{ieeetr}
\bibliography{bibliography/AI_Accelerators,
bibliography/github_repos,
bibliography/Inference_frameworks,
bibliography/LLM_HF_repos, 
bibliography/LLMs,
bibliography/Validation,
bibliography/related_extra,
bibliography/survey}

\listoffixmes

\end{document}